%% file: main.tex
\documentclass[graybox]{svmult}

\usepackage{type1cm}        % activate if the above 3 fonts are
\usepackage{makeidx}         % allows index generation
\usepackage{graphicx}        % standard LaTeX graphics tool
\usepackage{multicol}        % used for the two-column index
\usepackage[bottom]{footmisc}% places footnotes at page bottom

\usepackage{newtxtext}       % 
\usepackage{newtxmath}       % selects Times Roman as basic font

\usepackage{xcolor} % colored tables
\usepackage{subcaption}
\usepackage{graphicx}% Include figure files
\usepackage{dcolumn}% Align table columns on decimal point
\usepackage{bm}% bold math
\usepackage[flushleft]{threeparttable} % table footnote

\usepackage{dsfont}
\usepackage{color,psfrag}
\usepackage{mathtools,amscd}

\usepackage{soul}
\usepackage{xcolor} % colored tables

\usepackage{ifpdf}
    \ifpdf
\usepackage{tikz}
\usetikzlibrary{arrows,chains,matrix,positioning,scopes, fit, calc}
\usetikzlibrary{cd}
\usepackage{chemfig}
\fi
\usepackage{hyperref}

\hypersetup{
	colorlinks,
	linkcolor=blue,
	citecolor=blue, 
	filecolor=black,
	urlcolor=blue}

\makeindex             % used for the subject index
\begin{document}

\title*{Machine Learned Force Fields: Fundamentals, its reach, and challenges}
\author{Carlos A. Vital$^{a*}$, Rom\'an J. Armenta-Rico$^{a*}$, and Huziel E. Sauceda$^{a\dagger}$}
\institute{$\dagger$Huziel E. Sauceda, \email{huziel.sauceda@fisica.unam.mx}
\at $^a$Instituto de F\'isica, 
Universidad Nacional Aut\'onoma de M\'exico, Cd. de M\'exico C.P. 04510, Mexico. \\
$^*$These authors contributed equally.
}
\maketitle

\abstract{Highly accurate force fields are a mandatory requirement to generate predictive simulations. In this regard, Machine Learning Force Fields (MLFFs) have emerged as a revolutionary approach in computational chemistry and materials science, combining the accuracy of quantum mechanical methods with computational efficiency orders of magnitude superior to \textit{ab-initio} methods.
This chapter provides an introduction of the fundamentals of learning and how it is applied to construct MLFFs, detailing key methodologies such as neural network potentials and kernel-based models. 
Emphasis is placed on the construction of SchNet model, as one of the most elemental neral network-based force fields that are nowadays the basis of modern architectures. Additionally, the GDML framework is described in detail as an example of how the elegant formulation of kernel methods can be used to construct mathematically robust and physics-inspired MLFFs.
The ongoing advancements in MLFF development continue to expand their applicability, enabling precise simulations of large and complex systems that were previously beyond reach. This chapter concludes by highlighting the transformative impact of MLFFs on scientific research, underscoring their role in driving future discoveries in the fields of chemistry, physics, and materials science.}

%%%%%%%%%%%%%% GRAPHICAL ABSTRACT %%%%%%%%%%%%%
\begin{figure}
    \includegraphics[width=\textwidth]{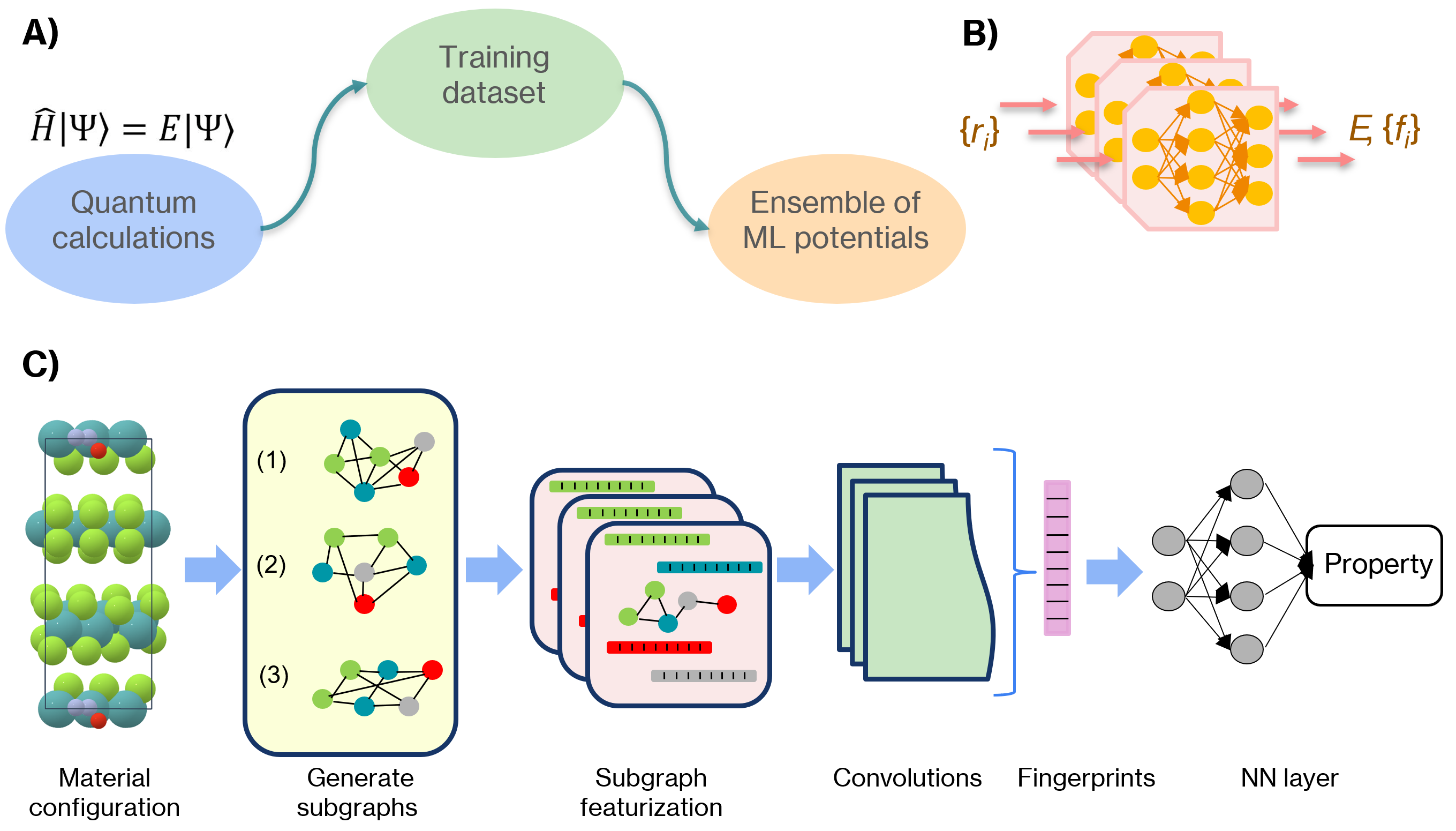}
    \caption{A) Machine learning starts from quantum calculations to learn material behavior through analytical potentials. B) Taking atomic coordinates, among other quantities as inputs, neural networks compute energies and learn atomic force fields as outputs. C) Common sub-processes for machine learning algorithms in material research. Mainly algorithms start with the codification of material configurations and continue with the learning through convolutional neural networks. And finally, multilayer perceptrons compute properties. }
    \label{fig:Graphical_abstract1}
\end{figure}
%%%%%%%%%%%%%%%%%%%%%%%%%%%%%%%%%%%%%%%%%%%%%%%

\section{Introduction}
\label{sec:Intro}

Accurate modeling of interatomic interactions is fundamental to understanding material properties and chemical processes at the atomic level. Based on fixed functional forms and empirical parameterization, traditional force fields have been extensively used in molecular dynamics (MD) and Monte Carlo (MC) simulations. However, these classical approaches often fail to accurately describe complex systems, particularly those involving bond breaking and bond formation, complex electronic interactions, and environments far from equilibrium. Quantum mechanical methods such as Density Functional Theory (DFT) or Coupled Clusters (CC) provide the necessary accuracy, but are computationally prohibitive for large systems and large-scale simulations. In this context, machine learning (ML)--based force fields have emerged as a transformative tool, bridging the gap between computational efficiency and quantum-level accuracy.

Machine learning-based force fields (MLFFs) are machine learning tools or methodologies assembled or structured to learn a specific function of the atomic coordinates: the potential energy surface (PES). These models are trained directly from quantum-mechanical calculations. Using neural networks, Gaussian processes, and other advanced ML techniques, these models can capture complex, high-dimensional relationships between atomic positions, energies, and forces without relying on predefined functional forms. One of the first successful models was the high-dimensional neural network architecture (HDNN), introduced by Behler and Parrinello~\cite{PES5}, a model capable of accurately modeling atomic interactions. This hybrid approach combined hand-crafted atomic descriptors with atomic multi-layer perceptrons. This foundational work paved the way for subsequent models, such as DeepMD Potential~\cite{zhang2018deep} and ANI~\cite{ANI2017} published ten years later. These modes improved accuracy, scalability, and transferability compared to HDNNs. Nevertheless, these approaches were working under the constraint of using predefined descriptors, meaning that a strong bias is imposed on the model and could hinder the expressibility of the networks to describe atomic interactions.

In contrast to this approach, Sch"utt \textit{ et al.}~\cite{SchNetNIPS2017,SchNet2018,schnetpack2018,schnetpack2023} introduced \textit{SchNet}, an innovative end-to-end learning framework based on message-passing neural networks (MPNN) that employ continuous filter convolutional layers to model quantum interactions. This new framework removed the need for hand-made descriptors and instead enabled learning the relevant atomic representation from the data. This revolutionized the research field in such a way that most modern MLFFs are based on their key contributions, convolutions and learned filters.
Soon after, it was realized that the invariant feature representations learned by SchNet could be generalized to equivariant representations. This marked the second generation of MPNN, which will be discussed in more detail in the text. 

In parallel to the neural network-based force field, MLFFs based on kernel methods were also being developed. Here, Gaussian Approximation Potential (GAP)~\cite{Bartok2010} and Gradient Domain Machine Learning (GDML)~\cite{gdml,sgdml} were the main models, each one with a very different take on the description of the atomic system. On the one hand, GAP takes the classical approach of representing the system's total energy as atomic contributions, while GDML uses a Hessian kernel to devise an analytically integrable force field to learn the total energy of the system without partitioning it into atomic contributions. Up to this day, GDML is the only global model published in the literature. 

In general, MLFFs have revolutionized atomistic simulations by significantly extending the range of accessible systems and time scales while maintaining near-quantum accuracy. Applications span from studying the dynamics of large biomolecules and nanomaterials to exploring new materials for energy storage and catalysis. Moreover, the ongoing integration of these force fields with active learning frameworks promises further advancements, enabling models to adaptively learn from new data and improve predictions in regions of chemical space that were previously unexplored.

However, challenges remain, particularly concerning the data requirements for training these models, the transferability across different chemical environments, and the interpretability of the learned representations. Addressing these issues is critical for the continued development and application of ML-based force fields in the field of computational chemistry and materials science.

In this chapter, we provide a comprehensive overview of the state-of-the-art MLFFs, discussing their underlying methodologies, key applications, and the challenges that lie ahead. We aim to highlight how these approaches are reshaping the landscape of atomistic simulations and driving innovation in the study of complex chemical systems.

% ==================== F I G U R E ===================
\begin{figure}[t]
\centering
\includegraphics[width=1.0\columnwidth]{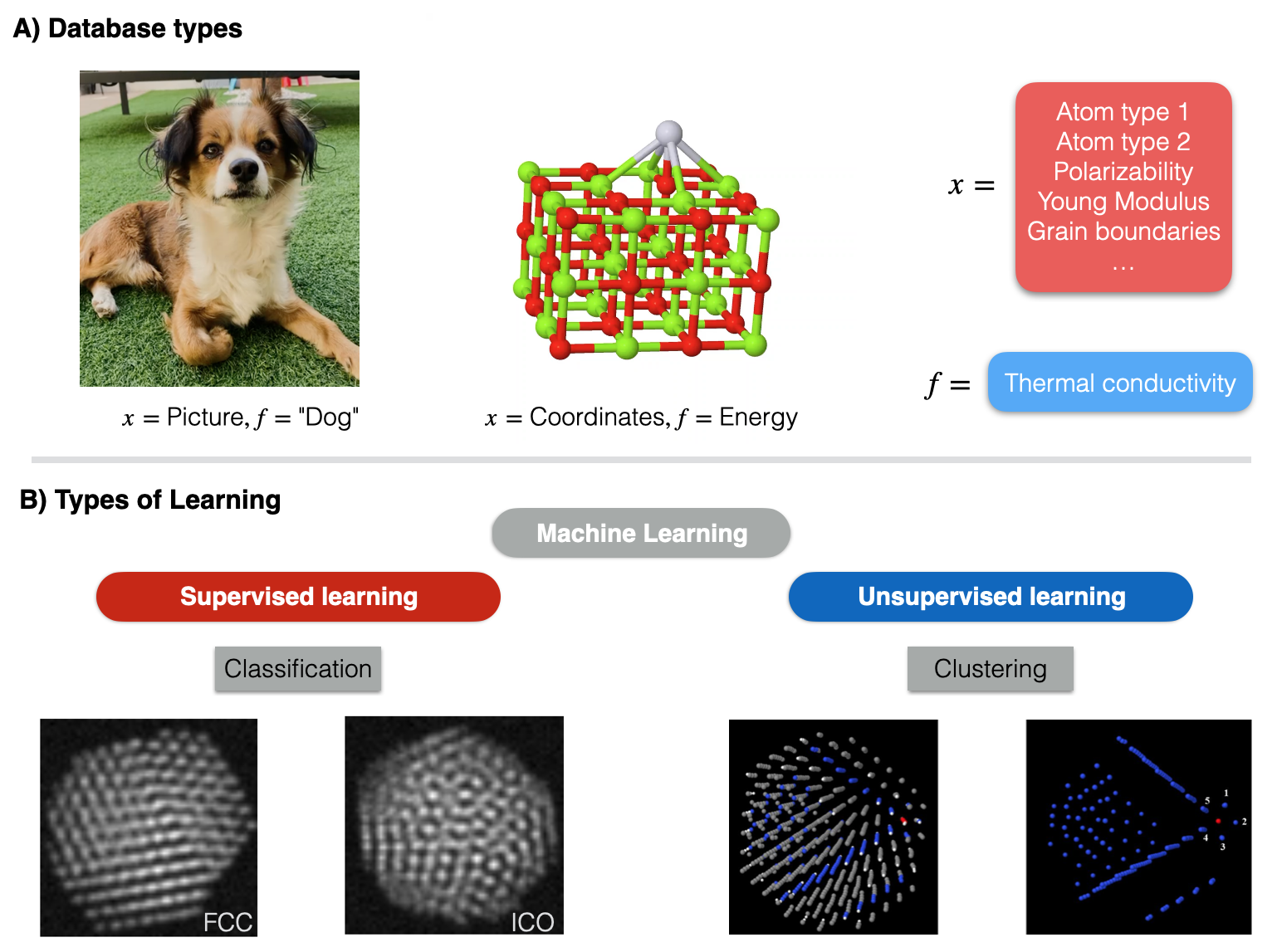}
\caption{A) Database types: Image classification, force field learning, and materials properties. B) Types of Learning. Supervised learning and unsupervised learning.}
\label{fig:learning}
\end{figure}
% =====================================================

% ======================================================
% =========================== Introduction to ML ========
% ======================================================
\section{Fundamentals of Machine Learning}
\label{sec:Funda} %
In this section, we briefly introduce a set of fundamental concepts applicable to any machine learning model. Let’s start with the database $D=\{f_j,x_j\}_{j=1}^M$ (See Fig.~\ref{fig:learning}). Here, $x_j$ is a feature vector or representation, for example, as shown in Fig.~\ref{fig:learning}-A, $x_j$ could be a picture, molecular or materials coordinates, or an array of physical properties, while the labels $f_j$ could be “dog”, energy, and the thermal conductivity, respectively. Something that is assumed is that there is a good amount of correlation or structure in the data, meaning that there is something to learn in the database. For example, the signal-to-noise ratio allows decoding the underlying structure of the signal.

There are mainly two types of learning, supervised and unsupervised. In simple terms, supervised learning refers to learning the correlation between the label $f$ and $x$ in the database, to construct a surrogate model or predictor $\hat f(x)$ either via a regression (e.g. learning a continuous function of $x$) or a classification problem (e.g. classification of different kind of flowers or structures from a Transmission Electron Microscope (TEM)) as shown in Fig.~\ref{fig:learning}-B. On the other hand, unsupervised learning focuses on analyzing the structure and correlations within $\{x_j\}_{j=1}^M$. A simple example is to find atoms with similar chemical environments in a material via Clustering or dimensionality reduction using Principal Component Analysis. Another branch of machine learning is Active Learning, which consists of dynamically or on-the-fly improving the performance of the models. In recent years, this has become very popular in different research fields, as we will see further in the text. As a side note, we would like to highlight that in the last couple of years, scientists have gone beyond the conventional ideas of Statistical Learning Theory and have used machine learning models as an ansatz for problems where a variational principle can be constructed or even as surrogate models for differential equations. In the next sections, we will elaborate on the topic.

\begin{figure}[t]
    \centering
    \includegraphics[width=1.0\columnwidth]{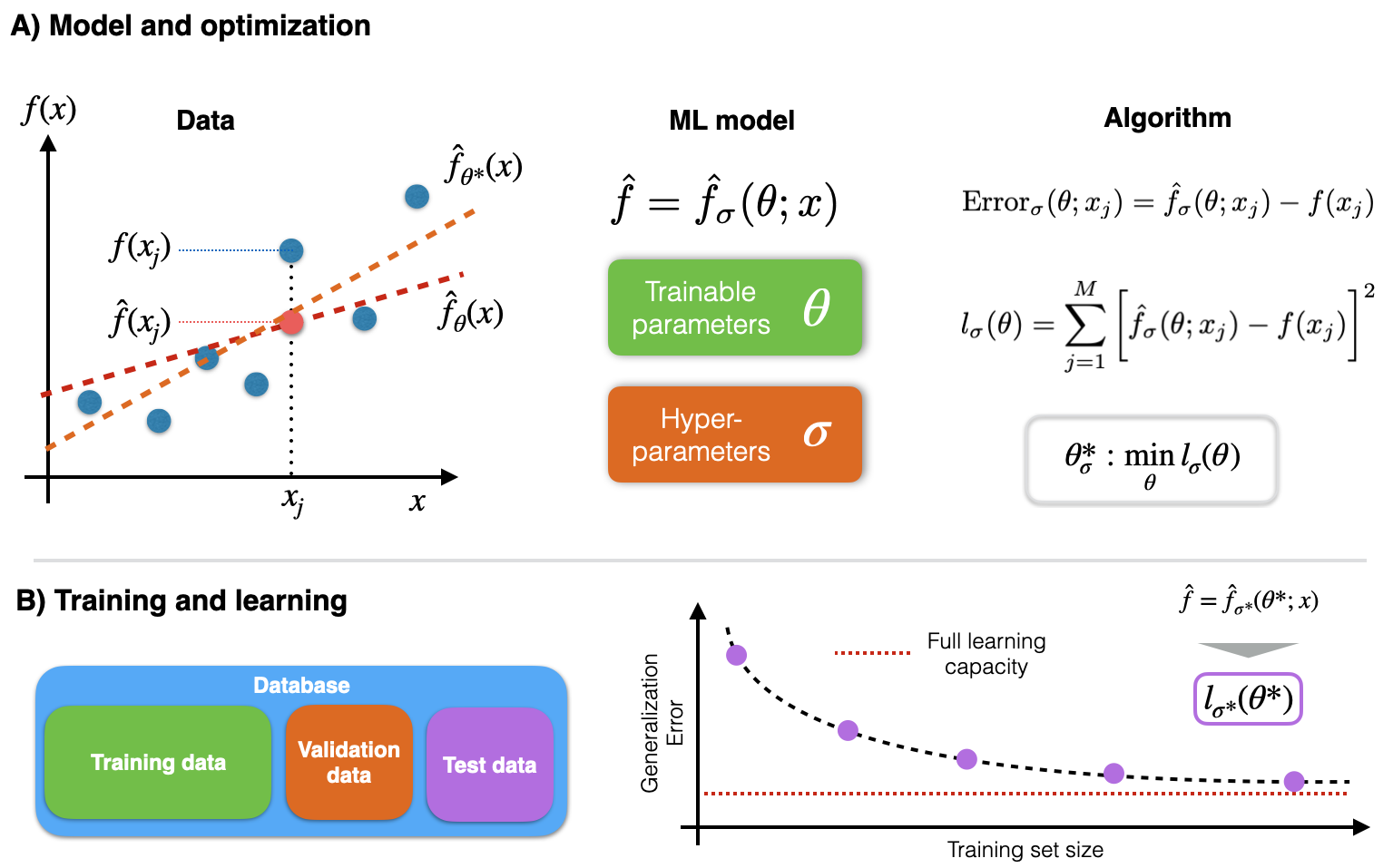}
    \caption{What does Learning mean? A) Model construction. B) Models training and generalization.}
    \label{fig:learning2}
\end{figure}

Once we have introduced the data structure and the types of learning; we focus on a more fundamental question: What does \textit{Learning} mean in a Machine Learning method? To answer this question, the first thing to do is take a look at the data. For the sake of simplicity, let’s consider the data in Fig.~\ref{fig:learning2}-A (left). It is worth highlighting that, the task at hand is not to fit a function to the data, and then analyze its functional behavior or dependencies as it is done in regular fitting, but instead, we want to create a surrogate model with an arbitrary functional form which can generate new predictions following the same patterns as encoded in the original training data. 

Based on the presented database, the next step is to select the functional of the model $\hat{f}_{\sigma} (\theta, x)$,  which can be, for example, a specific family of architectures like recurrent neural networks or graph neural networks, here $\theta$ are the trainable parameters and $\sigma$ are the hyperparameters, e.g. number of layers and activation functions. Here, we select our functional form to be a straight line.
Now, given the analytical form of the model, the next step is to train it. At this stage, another fundamental decision has to be made, how to measure the error. In this case, a regression problem, the $L_2$-norm provides good results. Then the \textit{loss function}, i.e. the error function to minimize, is given as in Fig.~\ref{fig:learning2}-B (right). Here, it is important to highlight that the database must be separated in three block, training, validation, and testing. Then, in the definition of the loss function $l_{\sigma}(\theta)$ only the training data is used. Consequently, to find the optimal value of $\theta$ only uses data from the green box in Fig.~\ref{fig:learning2}-B (left), while the search for the best set of hyperparameters is done using the unseen validation data. Once a set of optimal training $\theta^*$ and model-hyperparameters $\sigma^*$ is found, then the model's generalization error (test error) or reported model's accuracy is computed using the (unseen) test data. This allows us to know if the model learned, meaning that the model managed to decode the hidden correlations in the data, or just memorized the training data.

Another interesting behavior is the dependence of the test error on the amount of data used for training. Fig.~\ref{fig:learning2}-B (right) shows the typical learning curve, whereby increasing the amount of training data in the loss function $l_{\sigma}(\theta)$, the test error slowly decreases converging to a certain limit: the full learning capacity. This means that, at some point, the model will saturate its learning capacity reaching an error correlated to its flexibility or number of trainable parameters. Such an error will be larger than zero because the analytical form of the function used as a model is not the same function that generated the data.

Now, with this very compact introduction to learning theory, we describe some of the most popular machine learning methods used to construct MLFFs, neural networks, and kernel methods.

% ==================== F I G U R E ===================
\begin{figure}[t]
\centering
\includegraphics[width=1.0\columnwidth]{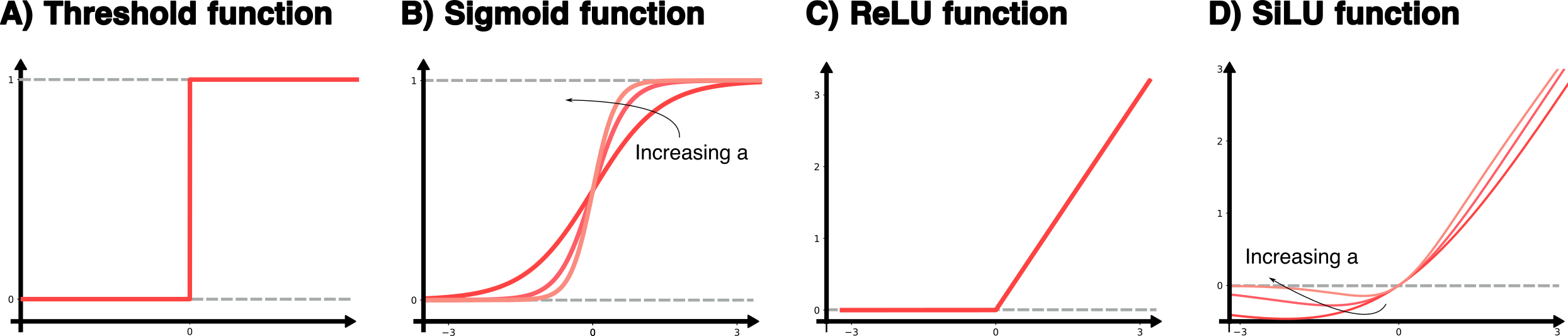}
\caption{Examples of activation functions. A) Threshold (or step) function. B) Sigmoid Function. C) ReLU function. D) SiLU function.}
\label{fig:ActivationFunctions}
\end{figure}
% =====================================================

%=================================
\section{Introduction to Neural Networks}
\label{sec:NN} %
%=================================
%
%=================================
\subsection{The perceptron}
\begin{figure}[t]
\centering
\includegraphics[width=1.0\columnwidth]{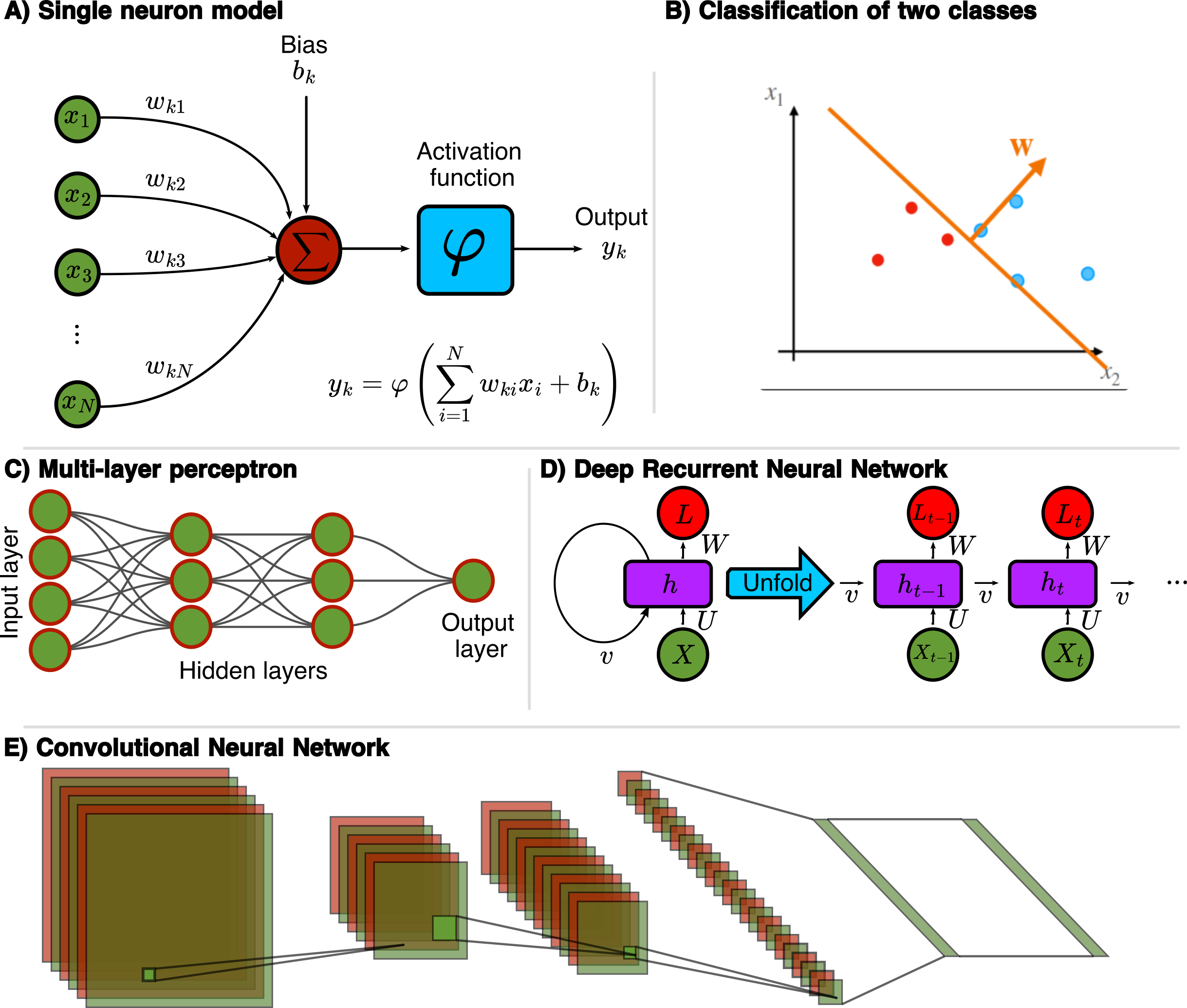}
\caption{A) Graphic view of a Neuron Model. B) Example of classification of two classes. C), D) and E) shows examples of neural network architectures. C) Multi-layer perceptron. D) Deep Recurrent Neural Network. E) Convolutional Neural Network.}
\label{fig:NeuralNetworks}
\end{figure}

%=================================
The perceptron is the simplest type of artificial neuron, introduced by Frank Rosenblatt in 1958~\cite{Rosenblatt1958} (See Fig.~\ref{fig:NeuralNetworks}-A). It is a basic information unit that mimics the behavior a biological neuron and forms the basis of more complex neural networks. A perceptron takes several input values $x_i$ ($\mathbf{x}\in\mathbb{R}^F$), applies weights $w_i$, sums them, adds a bias term $b$ and then passes the result through an activation function $\varphi$ to produce an Output $\in\mathbb{R}$:

 \begin{equation}
\text{Output}(\mathbf{x}) = \varphi\left(\sum_{i=1}^{F} w_i x_i + b\right).
\end{equation}

This model can be seen as a simple linear classifier when the activation function is a step function (Fig.~\ref{fig:ActivationFunctions}-A and -B), dividing input space into two domains (classes). Over the years, the activation function kept the name referencing its origins related to biological neurons, nevertheless, nowadays, \textit{non-linearity} is a much more accurate description because it highlights the critical role that these functions play in neural networks: introducing non-linear transformations that allow the network to learn complex patterns, i.e. they enable a neural network to model intricate functions by transforming the inputs at each neuron in a non-linear fashion to approximate complex mappings between inputs and outputs. Fig.~\ref{fig:ActivationFunctions} shows some popular nonlinearities. 

Activation functions have unique properties that define performance and efficiency during the training of a neural network. A clear example was the introduction of the ReLU (Rectified Linear Unit) non-linearity, $\sigma(x)=max(0,x)$. This is a widely used transformation in deep learning due to its simplicity and effectiveness in mitigating the vanishing gradient problem, a problem the limited applicability of neural network models while using traditional functions like the sigmoid function. 
Introducing ReLU was one of the great breakthroughs in deep learning, since it enabled the construction of very deep neural networks. It works by allowing only positive inputs to pass through while setting negative values to zero. This helps in faster convergence during training by keeping the gradient alive and avoiding saturation. However, training may suffer from inactive neurons or "dying neurons", where they stop contributing to the learning process. 
Traditional non-linearities, like \textit{Sigmoid} and \textit{Tanh}, still find some use in specific situations where a bounded output is necessary within hidden layers, despite their tendency to cause vanishing gradients. In general, \textit{Tanh} is preferred instead of sigmoid in regression problems because it results in faster convergence due to its zero-centered property.
Nowadays, one of the most popular non-linearities is \textit{SiLU} (Sigmoid Linear Unit), also known as the Swish activation function. It has gained popularity in regression learning and other machine learning tasks due to its smooth and non-monotonic behavior that combines the advantages of both ReLU and sigmoid functions: $\sigma(x)=x*Sigmoid(x)$. In particular, it has found great success in the latest \textit{equivariant networks}~\cite{unke2024e3x}.

Selecting the appropriate activation function goes beyond simply introducing non-linearity; it plays a crucial role in improving the learning dynamics, enhancing stability, and accelerating the convergence of the neural network.

%\subsection{Model of a neuron}

%A \textit{neuron} is the basic information--processing unit that composes all the neural network models, and it's composed by three basic elements:
%\begin{enumerate}
%\item A set of \textit{weights} that measures the importance of the value of a given neuron to the complete computation of the neural network. Specifically, a signal $x_j$ received by the neuron $k$ is multiplied by the synaptic weight $w_{kj}$, which can be real or complex. It also includes an externally applied bias, denoted by $b_k$.
%\item An adder for summing the input signals received by the neuron, weighted by the respective synaptic strengths.
%\item An \textit
%activation function that limits the permissible amplitude range of the output signal to some finite value. This activation function has to be non-linear to provide the complexity to the neural network model needed to solve complex problems.
%\end{enumerate}

% ==================== F I G U R E ===================

Something worth highlighting is that the perceptron is a \textit{universal approximator}, that is, this model converges in a finite number of steps to the correct classification of a data set, given that the training data is linearly separable. This theorem is called the \textit{Perceptron Convergence Theorem}~\cite{HaykinBook}. It is one of the most important theoretical foundations of machine learning, and its proof is the most celebrated result in this field.
Even though the perceptron convergence theorem leads to a universal approximator through a simple perceptron, this universality is limited to the classification of linearly separable patterns~\cite{HaykinBook,Efron1964}. To overcome the practical limitations of the simple perceptron, a more complex neural network is needed to solve more difficult tasks, such as classification of non-lineally separable data sets, clustering, associating, among others.

%=================================
\subsection{Multilayer perceptron}
%=================================

The Multilayer Perceptron (MLP) is one of the most fundamental types of artificial neural networks, being one of the go-to \textit{blocks} in MLFFs (See Fig.~\ref{fig:NeuralNetworks}-C).
An MLP consists of multiple fully connected layers of perception units or neurons, which introduces further complexity while adapting the dimensionality between input and output according to the specific use case.
In general, MLPs are small networks with only two or three layers, and are part of larger more intricate architectures, given that deep MLPs are extremely complicated to train due to, for example, vanishing or exploding gradients. For instance, the SchNet architecture uses a two-layer MLP to construct its filters, another two-layer MLP for atom-wise feature mixing. 
In a simple feed-forward neural network with one hidden layer, the output can be represented as,

\begin{equation}
\text{Output}(\mathbf{x}) = \varphi_2\left(\sum_{j=1}^{F_1} w_{2j} \varphi_1\left(\sum_{i=1}^{F} w_{1ij} x_i + b_{1j}\right) + b_2\right).
\end{equation}

This architecture expands the capabilities of a simple Rosenblatt's perceptron, since it offers the advantage of learning non-linear decision boundaries.

\subsection{The Architecture of a Neural Network}
% Hablar de capas densas, neuronas agrupadas en capas. Bullet points
Artificial neural networks are built with a set of single neurons connected in a given way. The definition of how neurons connect each other is what we call the \textit{architecture of a neural network}~\cite{HaykinBook,BishopBook}. 
Generally speaking, all the neurons in a neural networks are grouped in \textit{layers}, and these layers connect to other layers in a certain manner. The layer's hyperparameters define how the neurons are grouped, and connected, and how they pass information to other layers. 
One of the most commonly used layers are the so-called \textit{dense layers} (of \textit{fully connected layers}), where each neuron is connected to every neuron in the previous layer~\cite{GeronBook, Carleo2017}. This type of layer is often used in the initial layers of a neural network to learn low-level features from the input data. For more specific tasks, like image recognition, the \textit{convolutional layers}~\cite{karpathy2017} (See Fig.~\ref{fig:NeuralNetworks}-E) are typically used, since they apply convolution operations to the input data using learnable filters, enabling the network to capture spatial patterns and hierarchical features. On the other hand, \textit{recurrent layers}~\cite{Tealab2018} (See Fig.~\ref{fig:NeuralNetworks}-D) handle sequential data by maintaining an internal state that captures temporal dependencies; that is, the neurons in recurrent layers have a memory of the previous states of the data. These layers are used in tasks such as time series prediction, natural language processing, and speech recognition, among others.
The layers described are some of the most widely used types of layers in neural networks, but there are many more, each serving specific purposes and playing crucial roles in various types of deep learning architectures.

\subsection{Optimization algorithms}
\label{subsec:Optm}

Optimization methods are crucial for training neural networks and optimizing the parameters to minimize the loss function. The most used are \textit{gradient descent} \cite{Ruder2016}, which reduce the loss function by iteratively adjusting the model parameters in the opposite direction of the gradient of the loss function concerning the parameters. This method has a stochastic version, called \textit{stochastic gradient descent}, where the gradient is computed using a subset of the training data (mini-batches) rather than the entire dataset. It updates the parameters more frequently, leading to faster convergence and better generalization.
The \textit{Adaptive Gradient Algorithm} (Adagrad) adapts the learning rate for each parameter based on the historical gradients. It performs larger updates for infrequent parameters and smaller updates for frequent parameters, effectively handling sparse data. \textit{Roop Mean Square Propagation method} (RMSProp) is an adaptive learning rate optimization algorithm that divides the learning rate by exponentially decaying average of squared gradients. It helps to normalize the learning process and improve convergence, especially in the presence of noisy gradients. The \textit{Adaptive Moment Estimation} (Adam) is an adaptive learning rate optimization algorithm that combines the advantages of both AdaGrad and RMSProp. It maintains per-parameter learning rates and adapts them based on the first and second moments of the gradients. These optimization methods are crucial in training deep neural networks and are essential for achieving good performance and convergence properties in various machine learning tasks. 

% =============================================================
\section{Introduction to Kernel Methods}
\label{IntroKM}
% =============================================================
In this section, we briefly introduce the general ideas and concepts behind kernel methods, which are the base for some of the most popular MLFFs. Kernel methods are a powerful class of machine learning techniques that enable the modeling of complex, non-linear relationships within data by transforming the input space into a higher- or even infinite-dimensional feature space.
At the core of these methods lies the kernel function $\kappa$, which, in colloquial terms, is a measure of similarity between two vectors  $\Phi$ and $\Phi'$ living in a Hilbert space. Specifically, they are connected via $\kappa(\Phi,\Phi')=\langle \Phi,\Phi'\rangle$. Some of the key properties of kernels are: 1) \textit{Symmetry}. This property ensures that the similarity between two points remains consistent regardless of their order, $\kappa(\Phi,\Phi')=\kappa(\Phi',\Phi)$. 2) \textit{Positive Semi-Definiteness (PSD)}. This ensures that the corresponding kernel matrix has non-negative eigenvalues, an essential property for many optimization algorithms in machine learning, such as Gaussian Processes. This can be expressed mathematically as,

\begin{equation}
\sum_{i=1}^{n} \sum_{j=1}^{n} c_i c_j \kappa(\mathbf{x}_i, \mathbf{x}_j) \geq 0,
\end{equation}

with $c_j$ any set of coefficients. They have other important properties, such as linearity and translation invariance.

Some common kernel functions are the (1) Linear Kernel, $\kappa(\mathbf{x}, \mathbf{y}) = \mathbf{x} \cdot \mathbf{y}$, (2) Polynomial Kernel, $\kappa(\mathbf{x}, \mathbf{y}) = (\mathbf{x} \cdot \mathbf{y} + c)^d$, (3) Gaussian (RBF) Kernel, $\kappa(\mathbf{x}, \mathbf{y}) = \exp\left(-\frac{\|\mathbf{x} - \mathbf{y}\|^2}{2\sigma^2}\right)$, and finally a particularly important kernel in MLFFs, (4) the Mat\'ern Kernel,

\begin{equation}
\kappa(\mathbf{x}, \mathbf{y}) = \frac{2^{1-\nu}}{\Gamma(\nu)} \left(\frac{\sqrt{2\nu} \|\mathbf{x} - \mathbf{y}\|}{\rho}\right)^\nu K_\nu\left(\frac{\sqrt{2\nu} \|\mathbf{x} - \mathbf{y}\|}{\rho}\right)
\end{equation}

where, $\nu$ controls the smoothness of the function (larger values make the function smoother), $\rho$ is the length scale parameter, $\Gamma$ is the Gamma function and $K_{\nu}$ is the modified Bessel function of the second kind.

Now, to connect kernels with learning, we introduce the \textit{Representer Theorem}. This states that the optimal function $\hat f$ that minimizes a regularized risk minimization problem, as stated in statistical learning theory, has the form,

\begin{equation}
\hat f(x)=\sum_{l=1}^M \alpha_l \kappa(x,x_l)
\label{eq:krr_pred}
\end{equation}

where $\alpha_j$ are coefficients determined during the optimization process, $\kappa$ is a kernel function, and $x_j$ are training data points. The Representer Theorem applies to a wide range of learning problems, including classification, regression, and function approximation tasks involving kernel methods. In particular, is crucial in implementing Kernel Ridge Regression (KRR), which can perform function learning via non-linear regression models using kernel functions~\cite{BishopBook}.
KRR combines Ridge Regression (linear regression with $L_2$ regularization) with the \textit{kernel trick}, allowing it to learn non-linear relationships between input features $x$ and the target function $f$.
Now, the learning process consists in finding a \textit{predictor} or surrogate function $\hat f$ that maps inputs to outputs, minimizing the error between predictions and true values given a set of training data $\mathcal{S}=\{x_j,f_j=f(x_j)\}_{j=1}^M$. The objective of KRR is to minimize the regularized least squares loss function, 

\begin{equation}
    \min_{\alpha} \left\| \mathbf{K}\alpha - \mathbf{f} \right\|^2 + \lambda \alpha^\top \mathbf{K} \alpha
\label{eq:losskrr}
\end{equation}

where $\alpha$ is the vector of model coefficients in the kernel space, $\mathbf{f}$ is the vector of target values, $\lambda$ is the regularization parameter, and $\mathbf{K}$ is the kernel (Gram) matrix that captures the similarity between all pairs of training data points in the feature space,

\begin{equation}
\mathbf{K} = \begin{bmatrix}
\kappa(x_1, x_1) & \kappa(x_1, x_2) & \cdots & \kappa(x_1, x_M) \\
\kappa(x_2, x_1) & \kappa(x_2, x_2) & \cdots & \kappa(x_2, x_M) \\
\vdots & \vdots & \ddots & \vdots \\
\kappa(x_M, x_1) & \kappa(x_M, x_2) & \cdots & \kappa(x_M, x_M)
\end{bmatrix}.
\end{equation}

Equation~\ref{eq:losskrr} has an analytical solution via a the normal equation,

\begin{equation}
    \alpha = (\mathbf{K} + \lambda \mathbf{I})^{-1} \mathbf{f},
\label{eq:normal}
\end{equation}

with $\mathbf{I}$ the identity matrix, and $\lambda$ ensures that the matrix is invertible and controls regularization. Once the set of $\alpha$s is determined, we can use eq.~\ref{eq:krr_pred} to predict the value of the function $f$ for a new input feature $x$. This formula combines the learned coefficients with the similarity between the new input and each training sample. The validation and generalization error of this predictor is calculated using the methodology introduced in Section~\ref{sec:Funda}. As an extra comment in training KRR models, there are cases where solving eq.~\ref{eq:normal} analytically is not possible due to heavy problems with the rank of kernel matrix or simply because of computational limitations to invert the matrix (e.g. huge amounts of RAM required). In these cases, a deep-learning numerical optimizer could be used (see Section~\ref{subsec:Optm}), or in complicated scenarios, Chmiela's optimizer offer stable model training~\cite{ChmielaSciAdv2023}.

In general, the representer theorem is a cornerstone of kernel-based machine learning, providing the theoretical foundation that allows complex non-linear tasks to be addressed in an elegant and computationally efficient manner. Such a formal framework enabled the development of the robust Gradient Domain Machine Learning model described in Section~\ref{sec:gdml}.

% =============================================================
\section{Machine Learning in Chemical Reactions and Catalysis}
% =============================================================

\label{sec:MLCat}

\begin{figure}
    \includegraphics[width=\textwidth]{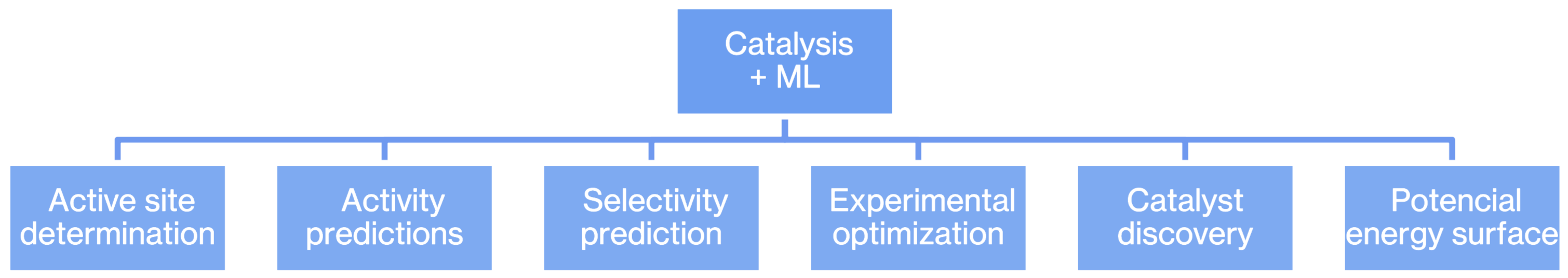}
    \caption{Some main applications of machine learning in catalysis, from active site determination to computing the potential energy surface of materials.}
    \label{fig:Graphical_abstract2}
\end{figure}

Before moving to more specific MLFF applications of ML, and introduced some concepts of ML, it is convenient to mention some of the applications of these tools in chemical reactions and catalysis.
%
%Ab initio methods continue supporting many \textit{in silico} studies of molecular systems. However, due to its computational cost, it is insufficient to integrate complexity as molecular systems increase in size and when their behavior depends strongly on the environment. Nevertheless, lots of work has been performed in an attempt to develop pure and applied knowledge \cite{Catalysis1,Catalysis2,Catalysis3,Catalysis4,Catalysis5, Catalysis6}.
%
In this context, machine learning has enabled tremendous growth in our capability to simulate, understand, and design new materials~\cite{Catalysis1,Catalysis2,Catalysis3,Catalysis4,Catalysis5, Catalysis6}. Nowadays, it is a fact that artificial intelligence has allowed everything from supporting experimental data to predicting certain structures with catalytic activity, as Fig.~\ref{fig:Graphical_abstract2} shows.
In general, ML tools learn from data to find insights to make fast predictions of target properties~\cite{ML_in_MS}. The present section mentions works in which ML methodologies were used in catalysis research.

\textit{Selectivity prediction}. 
Artificial neural networks have been used to predict product selectivity in a reaction, that is to say, product distribution considering these product components as the output layer of the neural network. Shigeharu Kite \textit{et al.} reported in 1994 the use of a neural network to estimate the product selectivity of styrene and various byproducts in the oxidative dehydrogenation of ethylbenzene on a series of promoted and unpromoted SnO$_2$ catalysts~\cite{KITE1994L173}.
They have used known properties of catalyst components as input data and observed catalytic performance as output data. The neural network adjusts iteratively its network pattern to make the calculated output data as close to the given output data as possible. The trained network pattern represents the relation between the given input and output data. Then, unknown output data, i.e., the catalytic performance to be obtained on a target catalyst, can be calculated by substituting the input data of the target catalyst in the network pattern thus trained. The input layer of the NN included parameters such as typical and unusual valence, ionic radius, the surface area of the catalyst, coordination number, and electronegativity, among others. The output layer includes the selectivities of the groups of reaction products in the oxidative dehydrogenation of ethylbenzene. Thereby, based on a set of promoted and unpromoted SnO$_2$ catalysts, the ANN can estimate the product distribution and, as shown in the Shigeharu's report, the accuracy of estimation is very satisfactory \cite{KITE1994L173}. Other works that use ML methodologies to compute selectivity are shown in \cite{Selectivity1, Selectivity2, Selectivity3}.

\textit{Catalyst design and discovery}
In general, computational predictions of catalyst structures, which thus far have been dominated by computationally expensive quantum mechanical methods such as density functional theory, are now being augmented by ML to accelerate the structure search of catalysts.
The purpose of catalyst design is to find the most efficient and suitable catalysts for a particular reaction. Although in the past, some researchers tried without success to design systematic catalysts, the trial and error and repetition of experiments were for a long time the main strategy in catalyst development. 
Hongliang \cite{Design1} discusses the discovery of perovskite oxides for use as air electrodes, achieved with machine learning. In the revision, stands out that Shuo Zhai et al \cite{Design2}, reported an experimentally validated machine-learning approach to accelerate the discovery of perovskite oxides for oxygen reduction in solid-oxide fuel cells. They curated a small dataset of perovskite oxides from the literature on which to train machine learning algorithms to learn underlying composition–activity correlations. For each material, they collated various features as metrics relating to the metal ions in the perovskites such as electronegativity, ionic radius, Lewis acidic strength, ionization energy, and tolerance factor (a predictor for the stability of perovskite structures). Machine learning algorithms were then employed to determine which showed the best generalizability in predicting the activity of materials \cite{Design2}.
The materials space for perovskite oxides is immense due to the composition and the phase, among others which influence catalytic activity. Then, the engineering of perovskite oxides needs inevitably accurate ML models that can predict new materials while providing interpretation or design rules for materials discovery ~\cite{Perovskitas_Discovery}. 
Musa et al \cite{Design3}, discuss some applications of ML to accelerate the search for catalyst structures. Other works that use ML methodologies to catalyst design and discovery materials are \cite{ML_design_discovery, Design4, discovery_failed_exp, Design5, Design6}.

\textit{Experimental condition optimizations}.
The classical method of reaction conditions optimization involves varying one parameter at a time that ignores the combined interactions between physicochemical parameters. However, it has significant drawbacks, such as requiring more experimental runs and time and failing to reveal the synergistic impact of processing parameters.
Barsi et al \cite{Optimization1}  performed an interesting work in which they studied the modeling and optimization of a reaction system to increase the efficiency of the process using ANN techniques. Particularly they studied an enzymatic reaction to produce wax esters, the enzymatic alcoholysis of vegetable oils. They used as parameters of the reaction conditions the temperature, the incubation time, the amount of enzyme, and the substrate molar ratio with which the yield of the reaction was optimized.
Thereby, they compare several neural-network architectures and topologies for the estimation and prediction of catalyzed synthesis of palm-based wax ester. Their comparison of predicted and experimental values revealed the ability of ANN in generalization for unknown data and implied that empirical models derived from ANN can be used to adequately describe the relationship between the input factors and output in their reaction. 
Some other works, in which the reaction conditions optimization is searched using machine learning methodologies, are \cite{Activity3,Optimization2, Optimization3,Optimization4, Optimization5}.

\textit{Active site determination}
In a heterogeneous catalyst, the catalytic activities are often dominated by a few specific surface sites, the active sites. Therefore designing active sites is the key to realizing high-performance heterogeneous catalysts \cite{Heterogeneous_catalyst}. Jinnouchi \textit{et al}. \cite{ActiveSite1} used an ML-based Bayesian linear regression method to predict the direct decomposition of NO on the Rh-Au alloy nanoparticles (NPs) using a local similarity kernel, which allows interrogation of catalytic activities based on local atomic configurations.  The proposed method was able to efficiently predict the energetics of catalytic reactions on nanoparticles using DFT data on single crystals (SC), and its combination with kinetic analysis was able to provide detailed information on structures of active sites and size (and composition) dependent catalytic activities.
The method relies on the fact that catalytic active sites are determined by their local atomic configurations; in other words, two sites having a similar atomic configuration should give similar catalytic activity. Based on that, the algorithm is composed of two steps: learning DFT data on SC surfaces and extrapolating the learning to NPs. The method showed that the kinetic analysis using the predicted energies can provide useful information of active sites. We can see other similar works based on ML methods to characterize active sites in \cite{ActiveSite2,ActiveSite3, ActiveSite4,ActiveSite5}.

\section{Overview and trends in MLFFs}
\label{sec:MLFF} %
%=============================================================
%
Machine Learning Force Fields (MLFFs) have emerged as a powerful tool in computational chemistry and materials science, enabling accurate and efficient simulations of atomistic systems with a wide range of complexities. 
Unlike traditional force fields, which rely on predefined functional forms and parameters, MLFFs learn directly from quantum mechanical data, capturing intricate interactions and enabling the study of large, complex systems that were previously computationally infeasible.
These methods leverage machine learning algorithms to learn the potential energy surfaces (PES) and interatomic forces, often achieving quantum chemical accuracy while maintaining computational efficiency. 
The development of MLFFs marks a significant advancement in atomistic modeling, allowing researchers to explore systems that were previously infeasible due to their size or the complexity of their interactions using either mechanistic/empirical force fields or \textit{ab-initio} simulations.

Traditional empirical force fields, often fail to accurately represent complex interactions, such as many-body effects and long-range correlations, because they rely on simplified, fixed functional forms. Quantum mechanical methods, such as density functional theory (DFT) or coupled-cluster calculations, provide high accuracy but are computationally expensive, and particularly for large systems, it becomes prohibitive to perform converged thermodynamic studies. Hence, MLFFs bridge this gap by learning from quantum chemical calculations, enabling accurate and scalable simulations~\cite{UnkeChemRev2021}.

%\subsection{Key Approaches in MLFF Development}
\textit{Neural Network based FF.}
Neural network-based force fields are among the most popular MLFFs due to their flexibility and accuracy. The Behler-Parrinello Neural Network model~\cite{PES5} was one of the first to introduce high-dimensional neural networks to predict atomic energies based on local atomic environments. Modern variants such as DeepMD~\cite{zhang2018deep} and ANI~\cite{ANI2017} advanced these models by incorporating complex architectures. Nevertheless, their use of hand-crafted descriptors hinders their flexibility and biases their functional forms. 
Contrasting this approach, SchNet~\cite{SchNet2018} introduced an end-to-end learning formalism that enables learning of atomic internal representations, a feature that most modern architectures follow.
Going further, PhysNet~\cite{UnkePhysNet2019} and SpookyNet~\cite{Unke2021SpookyNet} introduced explicit terms that account for long-range interactions such as electrostatics and van der Waals interactions. This addresses the intrinsic limitation of MLFF to only describe local interactions.

Later was realized that using atomic representation based only on interatomic distances was incomplete and that angular information was necessary to improve the accuracy of the models. In this direction, many equivariant neural networks have emerged, such as PaiNN~\cite{SchuttPaiNN2021}, NequIP~\cite{BatznerNequIP2022}, SpookyNet~\cite{Unke2021SpookyNet}, E3x~\cite{unke2024e3x}, and SO3krates~\cite{FrankSO3krates2024}.  
Equivariant neural networks incorporate physical symmetries directly into their learning processes. These models respect rotational, translational, group equivariant for representations, and permutational symmetries, enhancing both the accuracy and generalization of the force fields. By explicitly encoding these physical constraints, equivariant networks provide robust predictions that align closely with fundamental physical laws, making them ideal for applications in materials science and large-scale molecular simulations.

\textit{Kernel based FF.}
Using a different ML approach, some research groups have used kernel methods to reconstruct the PES of an atomistic system.
The first model proposed was the Gaussian Approximation Potentials (GAP) and later its associated descriptor Smooth Overlap of Atomic Positions (SOAP)~\cite{Bartok2015_GAP}, they are advanced tools in machine learning force fields. The combination of GAP and SOAP represents a significant advancement in the ability to model complex atomic interactions in a wide variety of material systems.
Another approach is the Gradient Domain Machine Learning (GDML) framework~\cite{gdml,sgdml,bigdml}. This is a powerful machine-learning approach designed to create accurate, data-efficient models of PES for molecular systems and materials. Unlike most MLFFs whose functional form focuses on an energy predictor and forces are obtained by differentiation, GDML directly learns the atomic forces, and its underlying PES is recovered by analytical integration. This allows the construction of global models that do not rely on arbitrary atomic energy partitioning. 
One of the key advantages of GDML is its ability to incorporate physical symmetries, such as rotational, translational, and permutational invariance, directly into the learning process. By encoding these symmetries in a symmetrized kernel function, GDML ensures that the predicted forces and energies adhere to fundamental physical principles, enhancing both accuracy and generalizability. 

%\subsection{Applications of MLFFs}
MLFFs have been applied successfully across a wide array of domains, including biomolecular dynamics, materials science, and the study of chemical reactions. 
They enable simulations of large systems with hundreds or thousands of atoms, which would be computationally prohibitive using traditional quantum mechanical methods. 
For instance, MLFFs have facilitated the study of protein folding, catalysis, and phase transitions with unprecedented accuracy, providing insights into rare events and transitions such as phase changes or complex reaction mechanisms~\cite{UnkeChemRev2021}.
One significant application of MLFFs is in path-integral molecular dynamics simulations, which are used to capture quantum effects in systems involving light atoms, such as hydrogen. This capability is crucial for accurately modeling hydrogen bonding, proton transfer, and other phenomena where nuclear quantum effects play a critical role~\cite{SchNet2018,ChmielaSciAdv2023,SaucedaNatCommun2021,sGDMLsoftware2019,sGDMLjcp}.

% Challenges and Future Directions
Despite their advantages, MLFFs face several challenges. One major issue is the need for large and diverse training datasets to ensure robust performance across different chemical environments. Ensuring transferability to systems that differ significantly from the training data is another ongoing area of research. 
Additionally, balancing computational efficiency with accuracy, and integrating physical constraints without compromising model flexibility, are critical considerations for the continued development of MLFFs.

Future research aims to address these challenges by developing more efficient training algorithms, integrating physical insights directly into model architectures, and exploring hybrid approaches that combine the strengths of different ML techniques. Enhancing the interpretability of MLFFs and quantifying their uncertainty in predictions are also important goals that will expand their applicability in scientific research and industrial applications.

%=============================================================
\section{Neural Network based Force Fields: The SchNet case}
\label{sec:schnet} %
%=============================================================
Nowadays, there is a plethora of MLFFs, and every week we have new models and architectures. Nevertheless, most of them share common grounds, they are based on "\textit{interaction}" blocks, i.e. message passing architectures (term coined by Gilmer \textit{et al.}~\cite{Gilmer2017}), and/or trainable filters and convolutions, concepts introduced in the DTNN~\cite{dtnn} and SchNet~\cite{SchNetNIPS2017,SchNet2018} architectures.
Hence, instead of explaining the latest architectures, we focus on presenting a clear introduction to the model SchNet, given that its features and inner workings are the base for modern architectures.

\begin{figure}
    \centering
    \includegraphics[width=1\linewidth]{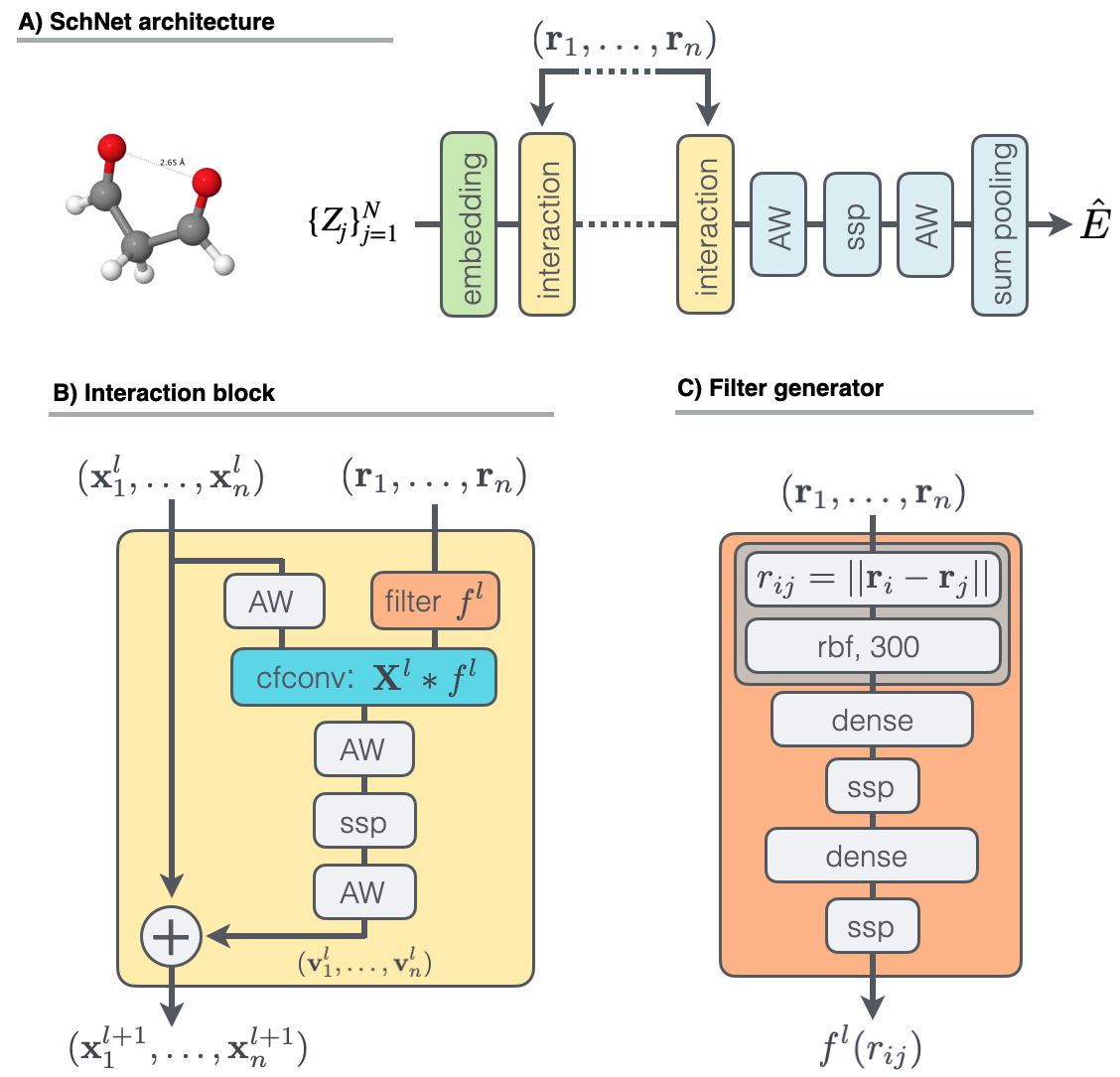}
    \caption{A) SchNet architecture. The atom embeddings are represented in a green box, interaction blocks in yellow, atom-wise (AW) networks, shifted-soft-plus activation functions (ssp), and sum poling operation are in blue boxes. B)
Interaction block architecture. The convolution layer is represented in cyan. C) Filter-generating network.}
    \label{fig:schnet}
\end{figure}

SchNet is a deep learning model specifically designed for predicting atomic properties, potential energy surfaces, and forces in molecular and materials systems. In the SchNet architecture, Sh\"utt \textit{et al.} introduced continuous-filter convolutional layers, tailored layers to model interactions between atoms based on their spatial arrangements according to correlations encoded in the data. 
The primary goal of SchNet is to learn atomic vector-feature representations from accurate quantum mechanical data, such as Density Functional Theory (DFT) calculations, without relying on predefined functional forms as in the original Behler-Parrinello networks. Then, these representations are used as inputs in a series of MLP networks to specialize the model to learn, for example, the PES or other chemical properties. Fig.~\ref{fig:schnet} shows the architecture and its different elements, which we describe in the next sections.

\subsection{Atom-type embeddings}
\begin{figure}
    \centering
    \includegraphics[width=1\linewidth]{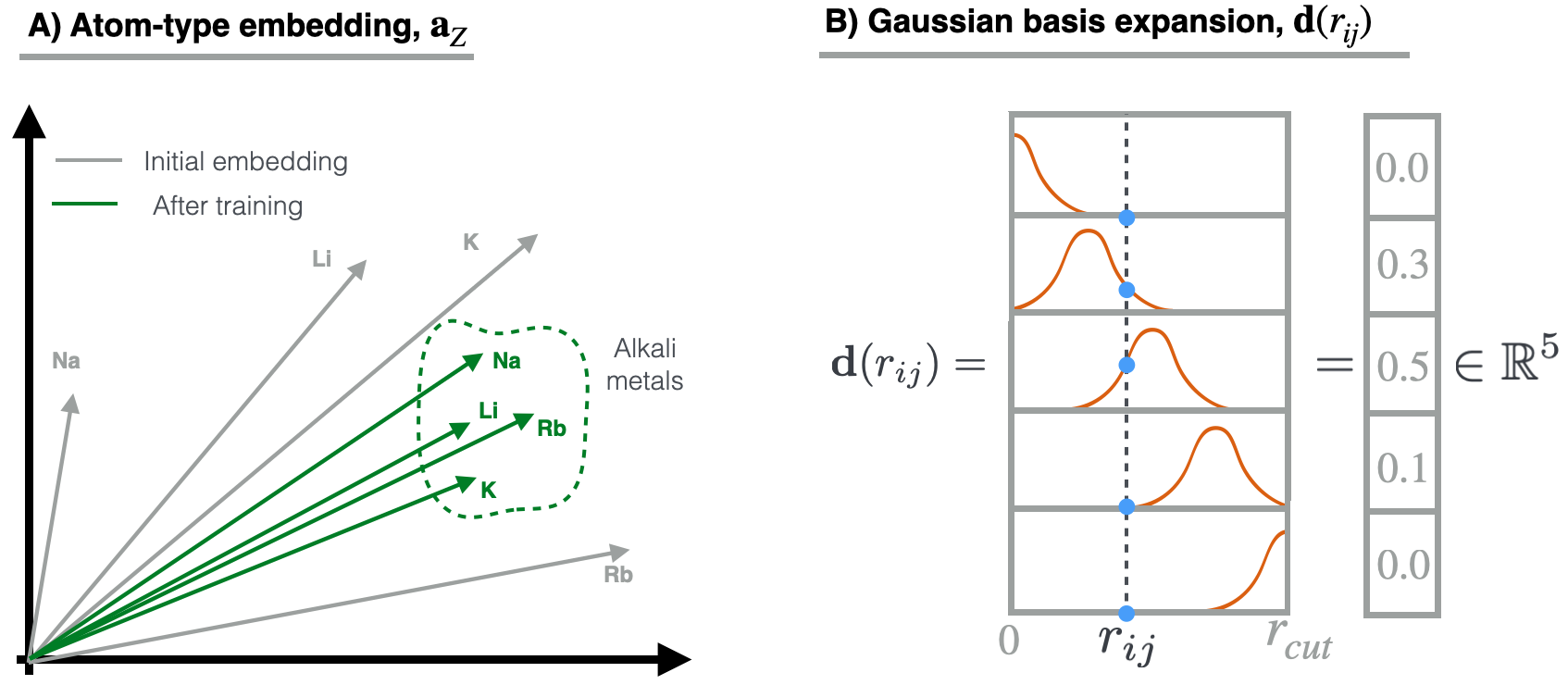}
    \caption{A) Atom-type embedding. A schematic representation of the randomly initialized vectors, and their clustering after training the networks. B) Gaussian basis expansion of the
interatomic distances. It shows an gaussian expansion in 5-dimensional space, where $r_{ij}$ is evaluated.}
    \label{fig:embedding}
\end{figure}

The initial transformation involves constructing embeddings for the N atoms in the systems, sorting them by chemical elements, $\{Z_j\}_{j=1}^N$. In this step, each atom's nuclear charge is mapped to a trainable, randomly initialized vector in a $F-$dimensional feature space: $T_{emb}: \mathbb{Z}^+ \to \mathbb{R}^F$. 
The initialization process involves sampling a vector for each atom type according to \(\mathbf{a}_Z \sim \mathcal{N}(0, 1/\sqrt{F})\). This atom-type embedding captures the quantum-chemical properties of a \textit{dressed atom}, meaning that it adapts based on the type of physical or chemical data the network is trained on. Consequently, the embeddings will group atoms with similar properties and spatially encode the correct atomic behavior (see Fig.~\ref{fig:embedding}-A).

The fundamental concept of the SchNet architecture is to iteratively transform and refine the initial atomic-type embedding $\mathbf{a}_Z$ until it encodes the chemical information hidden in the training data. After passing through $T$ interaction blocks the initial embedding transforms into $\mathbf{x}_Z^{(T)}$, where $\mathbf{x}_Z^{(0)}=\mathbf{a}_Z$.

\subsection{Interaction blocks}
SchNet introduced \textit{filter-generating networks}, a feature now common in many modern architectures. These networks modulate the atom-wise representations through linear transformations, effectively decoupling the learning of atomic embeddings from their geometric environment dependence.
Fig.~\ref{fig:schnet}-B shows the transformation of vector feature representation via \textit{atom-wise layers} and \textit{continuous-filter convolutions} (cfconv).
In this figure, \textit{Atom-wise (AW) layers} are dense layers that transform individual atomic representations $\mathbf{x}_i$ via $\Tilde{\mathbf{x}}_i = W\mathbf{x}_{i} + \mathbf{b}$, where weights W and biases $\mathbf{b}$ are shared across atoms. The main purpose of this layer is to mix information on the $F$ different feature dimensions of a single atom representation.

Another relevant part of the architecture is the f\textit{ilter-generating network} shown in Fig.~\ref{fig:schnet}-C. In general, this is a network that generates trainable radial functions (filters), $f(r_{ij})$. During training, these functions can capture interatomic interaction scales, but still, their interpretation is under debate.
The filter generator is constructed using two fully connected layers with shifted softplus (ssp) activation functions, taking as input a vector function $\mathbf{d}(r)\in \mathbb{R}^K$ (Gaussian expansion) that depends on the interatomic distances $r_{ij}$ (See Fig.~\ref{fig:embedding}-B). 
These filters introduce interatomic distance $r_{ij}$ dependence on the atomic feature representation $\mathbf{x}_j$.

Once the filter has been generated, the update rule for a given atom $i$ is the convolution layers given by,

\begin{equation}
    \Tilde{\mathbf{x}}_j=\mathbf{x}_j + \sum_{i\neq j}^{N} \mathbf{x}_i\circ f(||\mathbf{r}_j-\mathbf{r}_i||).
    \label{eq:cfconv}
\end{equation}

After this layer, further refinement is processed by a series of MLPs or AW layers (Fig.~\ref{fig:schnet}-B) to construct the message $\mathbf{v}_j$.

\subsection{The explicit The SchNet model for $T=2$}
In order to analytically expand the SchNet architecture, here we explicitly write the equation for the predictor energy $\hat E$ of a system with $N$ atoms and only \textit{two} interaction blocks.

For a system defined by the set tuples $\{(Z_j,\mathbf{r}_j)\}_{j=1}^N$, according to Fig.~\ref{fig:schnet}-A, the first operation is to generate the trainable embeddings $\mathbf{a}_{Z_j}=\mathbf{x}_j^0 \in \mathbb{R}^{32}$ for each atom-type, where we have chosen a feature space of 32 dimensions. If we have the aspirin molecule, then we will have the set $\{\mathbf{a}_{O},\mathbf{a}_{C},\mathbf{a}_{H}\}$ for oxygen, carbon, and hydrogen, respectively. Then, this initial feature representation is refined by exchanging information with atoms in its local environment via $T=2$ passes of the interaction block (Fig.~\ref{fig:schnet}-B). 
At the beginning, the atomic features $\mathbf{x}_j^0$ do not have geometrical information. Then the first transformation on the first pass of the interaction block to construct the message $\mathbf{v}_j^1$, $t=1$, is an AW (dense) layer, which mixes the 32 components of $\mathbf{x}_j^0$,

\begin{equation}
    {\Tilde{\mathbf{v}}}_j^{1} = W\mathbf{a}_{Z_j} + \mathbf{b}
\end{equation}

\noindent where $\Tilde{\mathbf{v}}_j^1$ represents the intermediate message state while processing $\mathbf{x}_j^0$ in the interaction block.
Next, the continuos-filter convolution is applied on the feature message vector $\Tilde{\mathbf{v}}_j^1$ with its environment. This is the first time geometric information is put into $\mathbf{x}_j$,

\begin{equation}
    \tilde{\mathbf{v}}_j^1 \mathrel{+}= \sum_{i\neq j}^{N} \mathbf{a}_{Z_i}\circ f^1(||\mathbf{r}_j-\mathbf{r}_i||).
    \label{eq:cfconv}
\end{equation}

After updating $\Tilde{\mathbf{v}}_j^1$ with cfconv with geometric radial information, its 32 entries are redistributed via an AW layer, then it is passed through a shifted softplus non-linearity and then mixed again. Thus, the message takes the form,

\begin{equation}
    \mathbf{v}_j^1= W'' \text{ssp}\bigg(W'\bigg[W\mathbf{a}_{Z_j} + \mathbf{b} + \sum_{i\neq j}^{N} \mathbf{a}_{Z_i}\circ f^1(||\mathbf{r}_j-\mathbf{r}_i||)\bigg] + \mathbf{b}'\bigg)+ \mathbf{b}''.
    \label{eq:interactionblock1}
\end{equation}

By explicitly showing the analytical form of the operations during the first pass of the interaction block, we see how the initial embedding $\mathbf{a}_{Z_j}$ is combined with other atomic embeddings $\mathbf{a}_{Z_i}$ modulated by their interatomic separation $r_{ij}$ via the radial filter. Lastly, the initial embedding $\mathbf{a}_{Z_j}$ is updated with the message $\mathbf{v}_{Z_j}^1$ which includes geometric information from its neighbors,

\begin{equation}
    \mathbf{x}_j^1=\mathbf{a}_{Z_j}+\mathbf{v}_j^1.
    \label{eq:resnet}
\end{equation}

This concludes the first pass of the interaction block.

Now, for the second interaction block, $t=2$, we only take the previous equation and change $\mathbf{a}_{Z_j}$ by $\mathbf{x}_{Z_j}^1$, hence the final expression for $\mathbf{x}_{Z_j}^2$ will be,

\begin{equation}
    \mathbf{x}_j^2=\mathbf{x}_j^1 + W'' \text{ssp}\bigg(W'\bigg[W\mathbf{x}_j^1 + \mathbf{b} + \sum_{i\neq j}^{N} \mathbf{x}_i^1\circ f^2(||\mathbf{r}_j-\mathbf{r}_i||)\bigg] + \mathbf{b}'\bigg)+ \mathbf{b}''.
    \label{eq:interactionblock1}
\end{equation}

If the neighborhood used to update the state $\mathbf{x}_j$ is defined by a cut-off radius $r_{cut}$, after $T$ interaction blocks, the final feature $\mathbf{x}_j^T$, in principle, it has access to the state of atoms located at a distance of up to $T*r_{cut}$. This is not entirely true, since information is attenuated due to information diffusion.

At this point, the atomic feature vectors are constructed, and now they can be used to train a specific quantity of interest, such has the total energy of the system. 
Then, in the original SchNet architecture, this is done by using the constructed atomic representations as inputs for a series of AW and ssp layers. Then, the total energy is approximated by the sum of atomic energies via,

\begin{equation}
    E_j= W'' \text{ssp}\bigg(W'\mathbf{x}_j^2 + \mathbf{b}'\bigg)+ \mathbf{b}''.
    \label{eq:Ei}
\end{equation}

the idea behind this is that the energy contribution of the atom $j$ to the total energy of the system can be constructed from its feature vector representation $\mathbf{x}_j^2$. Hence, the total energy is given by (\textit{sum pooling layer}),

\begin{equation}
    \hat E= \sum_{j=1}^N E_j.
    \label{eq:energy}
\end{equation}

Even thought, nowadays, there are much more convoluted architectures, the fundamental step by steps description presented here, is equivalent to those published in recent years.

\begin{figure}
    \centering
    \includegraphics[width=1\linewidth]{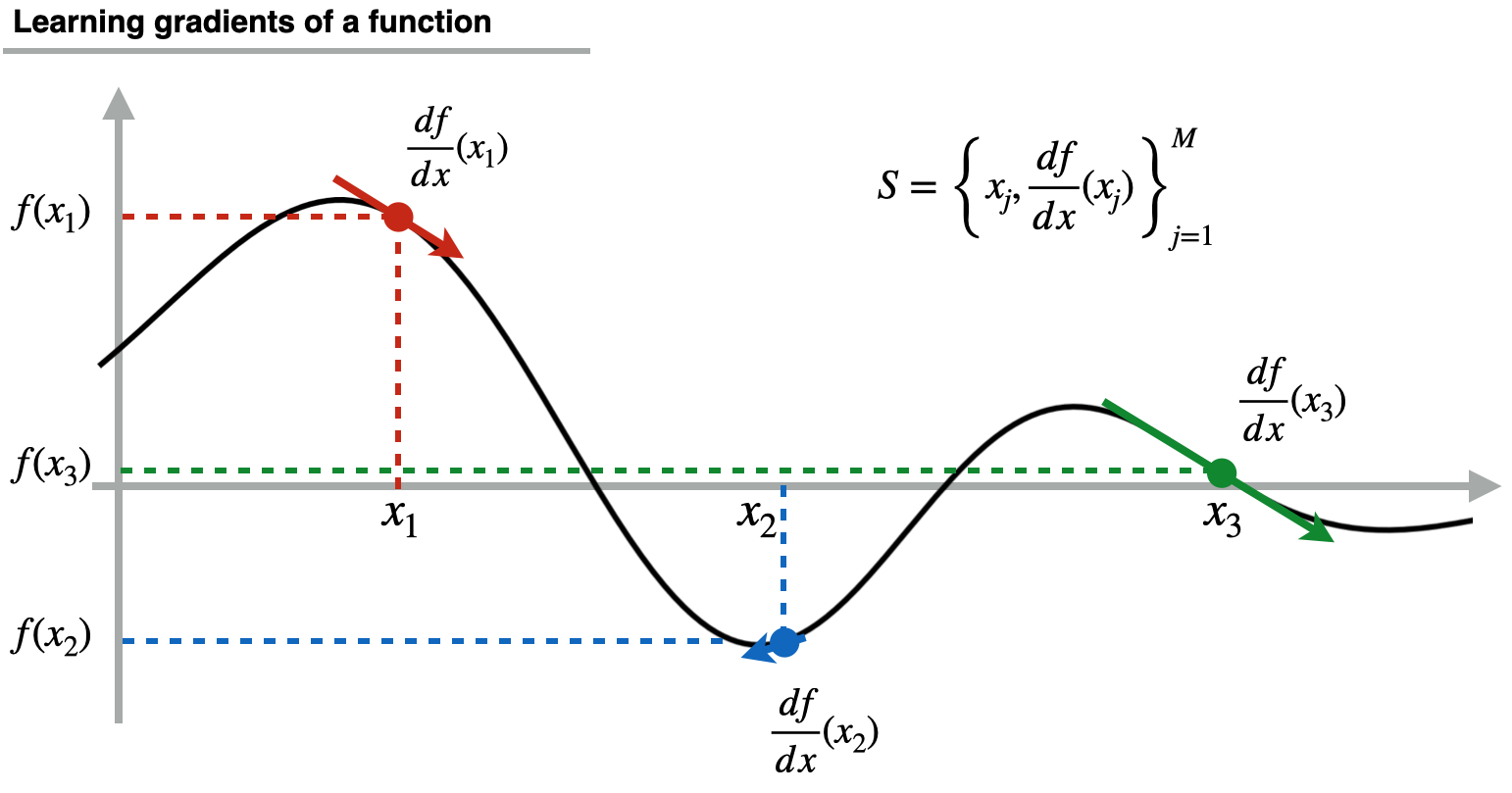}
    \caption{Database consisting in gradients fo a function $f(x)$.}
    \label{fig:datagrad}
\end{figure}

%=============================================================
\section{Kernel based Force Fields: The GDML framework}
\label{sec:gdml} %
%=============================================================
In Section~\ref{IntroKM} we introduced the concept of \textit{kernel} and how to use it for learning tasks via kernel ridge regression (KRR). It is possible to construct an MLFF model using simple KRR. The total energy can be defined as $\hat E(x)= \sum_l \alpha_l \kappa(x,x_l)$, eq.~\ref{eq:krr_pred}, and the atomic forces can be obtained by direct differentiation, $\mathbf{F}=-\nabla \hat E$. Nevertheless, this generates an MLFF that requires large amounts of data and still produces noisy atomic forces~\cite{gdml}. If we consider that databases such as MD17~\cite{gdml} and MD22~\cite{ChmielaSciAdv2023} are molecular dynamics trajectories that contain coordinates, energies and forces, and that training using energies imply that for each coordinates $\mathbf{x}_j$ we have a single label of energy $E_j$, while training using forces the label will contain $3N$ values. This means that training directly using atomic force labels is much more informative. Additionally, if we train a model that learns forces, we would like to also get the total energy directly by integrating the force field, since $\mathbf{F}=-\nabla E$.

We can redefine our learning problem by analyzing Fig.~\ref{fig:datagrad}, so we have a database consisting of the gradient of a function directly. In 1D, we could use again eq.~\ref{eq:krr_pred}, so we will have a predictor,

\begin{equation}
   \widehat{f'}(x) = \widehat{\frac{df}{dx}}(x) = \sum_l \alpha_l \kappa(x,x_l).
    \label{eq:krrgrad}
\end{equation}

We know that our data is de first derivative of a function, hence we could recover the underlying function $f$ by analytical integration,

\begin{equation}
   \widehat{f}(x) = \int_x dz \widehat{f'}(z) + c = \sum_l \alpha_l \int_x dz \kappa(z,x_l) + c.
    \label{eq:krrgradint}
\end{equation}

Now, this presents a big challenge, to recover the function $f$, we have to use an analytically integrable kernel. Instead of looking for an \textit{ad-hoc} kernel function with this property, since it can hinder the expressive power of the model, using the kernel properties to define a \textit{hessian kernel},

\begin{equation}
   k(x,y)\equiv \frac{\partial^2}{\partial x \partial y} \kappa(x,y),
    \label{eq:hessk}
\end{equation}

a kernel that by construction is integrable. Hence, if we use this kernel in eq.~\ref{eq:krrgradint} and consider radial kernel function, i.e. $\kappa(x,y)=\kappa(x-y)$, we obtain

\begin{equation}
   \widehat{f}(x) = \sum_l \alpha_l \int_x dz k(z,x_l) + c= \sum_l \alpha_l \frac{d}{d x} \kappa(x-x_l) + c.
    \label{eq:krrgradint}
\end{equation}

and the original KRR for $f'$ takes the form,

\begin{equation}
   \widehat{f'}(x) = \sum_l \alpha_l \frac{d^2}{d x^2} \kappa(x-x_l).
    \label{eq:gdml1D}
\end{equation}

From here, we can generalize these equations to the learning problem of a potential energy surface and atomic energies of a molecular system, and we get the Gradient Domain Machine Learning (GDML) framework~\cite{gdml}:

\begin{equation}
   \mathbf{F}(x) = \sum_l (\vec{\alpha}_l \cdot \nabla) \nabla \kappa(x-x_l),
    \label{eq:sgdml}
\end{equation}

\begin{equation}
   E(x) = -\sum_l (\vec{\alpha}_l \cdot \nabla) \kappa(x-x_l).
    \label{eq:sgdmle}
\end{equation}

This framework constitutes an elegant and descriptive model for constructing MLFFs.

After its introduction, and due to the mathematical properties of kernel functions, it was realized that GDML could be symmetrized by the symmetry group of the system under study, either a molecule or a material with periodic boundary conditions. Here, the only difference was which symmetry group $G$ to use to symmetrize the kernel function. By definition, the sum of two or more kernels is still a kernel, then a symmetric kernel is constructed by adding all relevant symmetric permutations $\mathbf{P}$ in a system,

\begin{equation}
   \kappa_{sym}(x,y) = \sum_{\mathbf{P}\in G} \kappa(x-\mathbf{P}x_l).
    \label{eq:symK}
\end{equation}

Using the symmetric kernel for molecules in the GDML framework resulted in the symmetric-Gradient Domain Machine Learning (sGDML) model~\cite{sgdml}. Furthermore, by using eq.~\ref{eq:symK} with the translational group in solids together with the Bravais group of the lattice, enabled the construction of the Bravais-Inspired GDML (BIGDML) model~\cite{bigdml}. Both models have proved to achieve extreme accuracies using only a handful of training data points.

\section{Summary and Concluding Remarks}
\label{sec:summary} %
Machine Learning Force Fields represent a transformative advancement in computational chemistry and materials science, offering a powerful alternative to traditional force fields and quantum mechanical methods. By leveraging machine learning techniques, such as neural networks, Gaussian processes, and kernel methods, MLFFs are capable of capturing complex interatomic interactions with near-quantum accuracy while maintaining the computational efficiency required for large-scale simulations. These models have proven to be highly adaptable, handling diverse chemical environments, including molecules, materials, and interfaces, and providing valuable insights into reaction mechanisms, phase transitions, and molecular dynamics over extended time scales.

A key strength of MLFFs lies in their ability to incorporate physical principles, such as symmetry and conservation laws, directly into their learning processes, which enhances both the accuracy and generalizability of the models. Techniques like equivariant neural networks and advanced descriptors such as SOAP have further improved the fidelity of these models, enabling them to respect fundamental physical constraints. Additionally, approaches like the GDML framework has demonstrated that highly accurate models can be constructed with relatively small training datasets, offering significant improvements in data efficiency compared to earlier methods.

However, the practical application of MLFFs is not without challenges. Ensuring the transferability of models to new chemical spaces, balancing computational efficiency with accuracy, and effectively managing uncertainty remain ongoing areas of research. Furthermore, integrating MLFFs into broader workflows for automated discovery and simulation presents additional opportunities for innovation. Despite these challenges, the rapid progress in the development of MLFFs continues to expand their impact, providing an indispensable tool for exploring the atomic-level behavior of complex systems with unprecedented precision.

In conclusion, MLFFs are reshaping the landscape of atomistic modeling, allowing researchers to overcome the limitations of traditional methods and enabling the detailed exploration of molecular and material systems. As the field continues to evolve, further advancements in model architectures, data efficiency, and interpretability are expected to drive even greater adoption of MLFFs across scientific disciplines. By combining the accuracy of quantum mechanical calculations with the speed of machine learning, MLFFs are not only enhancing our understanding of atomic interactions but also paving the way for new discoveries in chemistry, physics, and materials science.

\section{Acknowledgments}
C.A.V. and R.J.A.R. acknowledge CONAHCyT for the scholarship provided for being part of Programa de Doctorado en Ciencia e Ingenieria de Materiales and Programa de Doctorado en Ciencias (Física) at UNAM, respectively.
H.E.S. acknowledges support from DGTIC-UNAM under Project LANCAD-UNAM-DGTIC-419 and from Grant UNAM DGAPA PAPIIT No. IA106023, and CONAHCyT project CF-2023-I-468. 
Also, H.E.S acknowledges the financial support of PIIF 2023.
%Also, H.E.S acknowledges Carlos Ernesto L\'opez Natar\'en for helping with the high-performance computing infrastructure.
Correspondence should be addressed to H.E.S.

\input{referenc}

%\bibliographystyle{abbrv}
%\bibliography{cites}

\end{document}